\documentclass[twocolumn,aps,prc,superscriptaddress]{revtex4-2}
\usepackage{amssymb}
\usepackage{amsmath,bm}
\usepackage{graphicx}
\usepackage[normalem]{ulem}
\usepackage{multirow}
\usepackage[colorlinks,linkcolor=blue,urlcolor=blue,citecolor=blue]{hyperref}
\usepackage{lipsum}
\usepackage[usenames, dvipsnames]{xcolor}
\usepackage{tensor}
\usepackage{isotope}
\usepackage{amsmath}
\allowdisplaybreaks[4]
 
\renewcommand{\sout}{\bgroup \color{red} \ULdepth=-.5ex \ULset}

\begin{document}

\title{Relativistic kinetic approach to light nuclei production in high-energy nuclear collisions}

\author{Kai-Jia Sun}
\email{kjsun@tamu.edu}
\affiliation{Cyclotron Institute and Department of Physics and Astronomy, Texas A\&M University, College Station, Texas 77843, USA}

\author{Rui Wang}
\email{wangrui@sinap.ac.cn}
\thanks{\\Kai-Jia Sun and Rui Wang contributed equally to this work.}
\affiliation{Key Laboratory of Nuclear Physics and Ion-beam Application~(MOE), Institute of Modern Physics, Fudan University, Shanghai $200433$, China}
\affiliation{Shanghai Institute of Applied Physics, Chinese Academy of Sciences, Shanghai $201800$, China}

\author{Che Ming Ko}
\email{ko@comp.tamu.edu}
\affiliation{Cyclotron Institute and Department of Physics and Astronomy, Texas A\&M University, College Station, Texas 77843, USA}

\author{Yu-Gang Ma}
\email{mayugang@fudan.edu.cn}
\affiliation{Key Laboratory of Nuclear Physics and Ion-beam Application~(MOE), Institute of Modern Physics, Fudan University, Shanghai $200433$, China}
\affiliation{Shanghai Institute of Applied Physics, Chinese Academy of Sciences, Shanghai $201800$, China}

\author{Chun Shen}
\email{chunshen@wayne.edu}
\affiliation{Department of Physics and Astronomy, Wayne State University, Detroit, MI 48201, USA}
\affiliation{RIKEN BNL Research Center, Brookhaven National Laboratory, Upton, NY 11973, USA}

\date{\today}
\begin{abstract} 
Understanding the production mechanism of light (anti-)nuclei  in high-energy   nuclear collisions and cosmic rays  has been a  long-standing   problem in nuclear physics.  In the present study, we develop a stochastic method to solve the relativistic  kinetic equations for light nuclei production from many-body reactions with the inclusion of their finite sizes.
The  present approach  gives an excellent description of the deuteron and helium-3  data from central Au+Au (Pb+Pb) collisions at $\sqrt{s_{\rm NN}}$ $=$ $200~\rm GeV$ ($2.76~\rm TeV$). It can also naturally explain their suppressed production in $pp$ collisions at 7 TeV as a result of their finite sizes.
%cluster e.g., deuteron ($d$) and triton ($t$) through many-body scatterings of pion~($\pi$) and nucleon~($N$), e.g., $\pi NN\leftrightarrow \pi d$, $\pi Nd\leftrightarrow \pi t$, and $\pi NNN\leftrightarrow \pi t$.   
\end{abstract}

\maketitle

\section{Introduction}
Besides   ordinary  hadrons in the Standard Model,   light   nuclei, such as deuteron ($d$), helium-3 ($^3\text{He}$), triton ($t$), helium-4 ($^4\text{He}$), hypertriton ($^3_\Lambda \text{H}$) and their antiparticles, have also been observed in high energy nucleus-nucleus ($AA$), proton-nucleus ($pA$), and $pp$ collisions at  the Relativistic Heavy Ion Collider (RHIC) and the   Large Hadron  Collider (LHC)~\cite{STARSc328,STARNt473,STARNtP16}.  Although rarely produced, these weakly bound nuclei   carry important information on the space-time geometry~\cite{MroPLB248}, entropy production~\cite{CsePR131}, and   the  QCD phase transitions~\cite{SKJPLB781,SKJPLB816,ShuEPJA56,ZDICNNC13} in these  collisions. They can also provide   indirect information   on the  dark matter  interaction  in   cosmic rays due to   their ultra-low astrophysical background~\cite{BluPRD96,KouICHEP39,DoeJCAP2020}.

Different mechanisms have been proposed to describe light nuclei production in   nuclear reactions.  For heavy-ion collisions at high energies, the created   fireball evolves   through stages of  pre-equilibrium, the Quark-Gluon Plasma (QGP) or partonic expansion, the hadronization, and  subsequent hadronic scatterings and decays until kinetic freeze-out.  In the statistical hadronization model~\cite{AndNt561},   light nuclei are thermally produced during the hadronization of QGP and  their yields are assumed to remain unchanged  during the hadronic evolution.   In the coalescence model~\cite{SchPRC59,Bellini:2020cbj}, light nuclei are produced from nucleons close  in phase space at  the kinetic freeze-out.  Since  the cross section  for the deuteron dissociation by pion reaction $\pi d \rightarrow \pi NN$ is  about $100~\rm{mb}$~\cite{PDG2020}, the  rate for the  inverse process $\pi NN \rightarrow \pi d$   in hadronic matter is also large due to the time-reversal symmetry. Therefore,  to understand light nuclei production in high-energy nuclear collisions, it is essential to include their formation and   dissociation  dynamics  in these collisions.     The process $\pi d \leftrightarrow \pi NN$ has   recently been studied in Ref.~\cite{OliPRC99}   by approximating it as   two-step processes of $NN\leftrightarrow d'$ and $\pi d' \leftrightarrow \pi d$ with $d'$ being a fictitious dibaryon resonance. For nuclear reactions at low energies, the process $N NN\leftrightarrow N d$ is more important because the produced matter consists mainly of baryons~\cite{DanNPA533}.  In all previous transport model studies~\cite{DanNPA533,OhPRC76,OhPRC80,OliPRC99}, light nuclei have been considered as point particles. This  is a valid assumption in  large collision systems, but no longer holds  in small systems  like   $pp$ collisions,  in which the size of produced matter is comparable to or even smaller than those of light   nuclei such as the $d$ and $^3_\Lambda \text{H}$~\cite{SKJPLB792,ZW2105}. Other proposed mechanisms for light nuclei production in high-energy nuclear collisions include the decay of a compact quark droplet~\cite{BraPR621} or pre-formed nuclear clusters~\cite{ShuPRC100}.  In spite of these efforts, the production mechanism of  light (anti-)(hyper-)nuclei  in nuclear collisions has not yet been
unambiguously understood~\cite{CJHPR760,Ono:2019jxm,DonEPJA56}.

\begin{figure}[t]
  \centering
  \includegraphics[width=6.5cm]{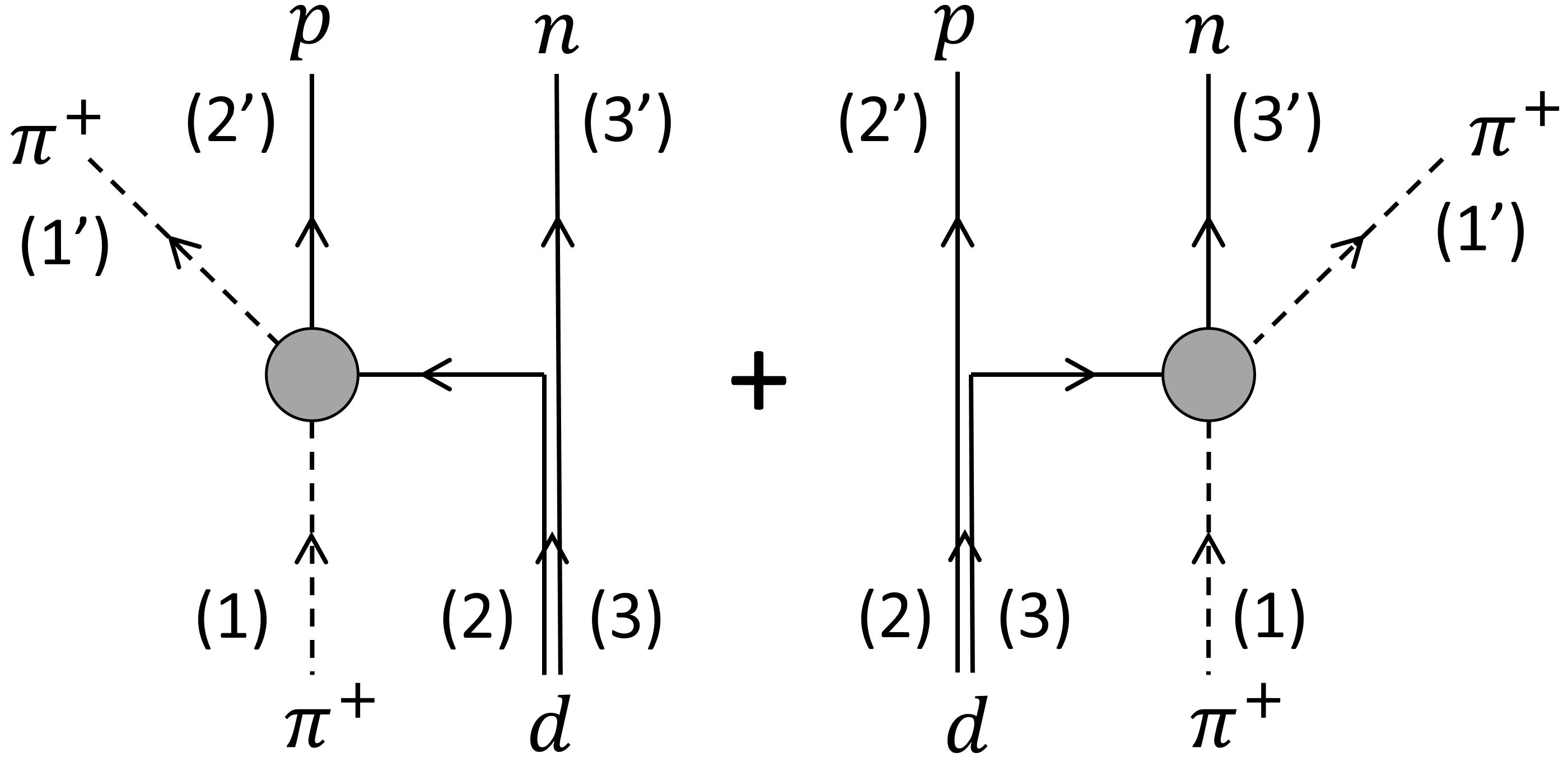}
  \caption{  Diagrams for the reaction  $\pi^+d\leftrightarrow \pi^+np$   in  the impulse approximation. Inside the bubble is a $\Delta$ resonance   as  an intermediate state. }
  \label{pic:piond}
\end{figure}

In the present article, we study the dynamics of light  nuclei in nuclear collisions by the relativistic kinetic equations that were  derived from the time evolution of  the system's Green's functions defined on a closed path in time~\cite{DanAP152,BotPR198,DanNPA533,DanPLB274}.   In this approach, the effect of finite sizes of light nuclei, which is usually neglected, can be included by retaining the relative coordinates  between nucleons in the gradient expansion of the Green's function~\cite{DanAP152,BotPR198}. A central quantity in the  relativistic kinetic equations is the many-body transition matrix  for light nuclei production and dissociation, which  will be evaluated here by using  the impulse approximation.    To illustrate the validity of the impulse approximation, we take $\pi d\to \pi NN$ as an example.  Since the  typical temperature of the hadronic matter in high-energy nuclear collisions is  100-150 MeV,  the pion thermal wavelength  is then around  0.4-0.5 fm  and  is  much smaller than the deuteron size of  about 3 fm.
A pion thus has a sufficiently large momentum to resolve the two constituent nucleons in the deuteron, resulting its scattering by one nucleon  with the other nucleon being quasifree, as shown in Fig.~\ref{pic:piond}. Therefore, the transition amplitude $\mathcal{M}_{\pi d\rightarrow \pi NN}$ for this reaction can be approximately factorized and is proportional to $\langle \tilde{p}_N|\phi_d \rangle\mathcal{M}_{\pi N\rightarrow \pi N}$, with $\langle \tilde{p}_N|\phi_d \rangle$ and $\mathcal{M}_{\pi N\rightarrow \pi N}$ being the deuteron wave function and the $\pi$-$N$ elastic scattering amplitude, respectively.  The impulse  or quasifree approximation was previously used  in studying  deuteron  dissociation in low-energy heavy-ion collisions~\cite{ChePR80, DanNPA533} and also  $J/\Psi$  dissociation by partons in relativistic heavy-ion collisions~\cite{GraPLB523}.
Under this approximation, the inverse   reaction $\pi NN\to\pi d$ can be viewed as a superposition of two subprocesses, the scattering between a pion and a nucleon with the final-state nucleon sightly off mass shell, and followed by the fusion or coalescence of the off-shell nucleon and another nucleon to form a deuteron.  The impulse approximation can be generalized straightforwardly to study the production and dissociation of heavier nuclei that involve more than two nucleons.

%  from the processes   $\pi Nd\leftrightarrow \pi t$ and $\pi NNN\leftrightarrow \pi t$.

\section{Relativistic  kinetic equation}
We start by considering one specific channel $\pi^+d\leftrightarrow \pi^+np$. Its effect on   deuteron production and dissociation in nuclear collisions can be described by the relativistic   kinetic equation~\cite{BotPR198,DanNPA533}, 
\begin{eqnarray}
\frac{\partial f_d}{\partial t_d}+\frac{  {\bf P}}{\  {E_d}}\cdot \frac{\partial f_d}{\partial {\bf R}}=-\mathcal{K}^{>}f_d + \mathcal{K}^{<}(1+f_d). \label{eq:deu_kin}
\end{eqnarray}
where the deuteron distribution function $f_d({\bf R},{\bf P})$ is normalized as $g_d(2\pi)^{-3}\int \text{d}^3{\bf R}\text{d}^3{\bf P}f_d=N_d$ with $g_d=3$ and $N_d$ being the deuteron statistical spin factor and total yield, respectively. 

On the l.h.s. of Eq.~(\ref{eq:deu_kin}), which denotes the drift term, we have treated deuteron as a free particle by neglecting its interaction with medium. The two terms on the r.h.s. of Eq.~(\ref{eq:deu_kin}) describe{s} the deuteron dissociation and production rates, respectively. Under the impulse approximation, the collision integral on r.h.s. is given by~\cite{DanNPA533,DanPLB274} 
 \begin{eqnarray}
&&\frac{1}{2g_dE_d}\int \prod_{i=1',3'}\frac{\text{d}^3{\bf p}_{i}}{(2\pi)^32E_{i}} \frac{\text{d}^3{\bf p}_{\pi}}{(2\pi)^32E_{\pi}} \frac{E_d\text{d}^3{\bf r}}{m_d}  \notag \\
&&\times    {2m_dW_d(\tilde{\bf r},\tilde{\bf p})\big{(}\overline{|\mathcal{M}_{\pi^+ n\rightarrow \pi^+ n}|^2} + n\leftrightarrow p\big{)}}  \notag\\
&&\times \Big[ -\big{(}\prod_{i=1'}^{3'} (1\pm f_i)\big{)}g_\pi f_\pi g_df_d +\frac{3}{4}\big{(}\prod_{i=1'}^{3'} g_if_{i}\big{)} \notag\\
&&~~~~\times (1+f_\pi)(1+f_d)\Big]\times  (2\pi)^4\delta^4(p_\text{in}-p_\text{out}), \label{eq:deu_rate}
\end{eqnarray}
from which ${\mathcal K}^>$ and ${\mathcal K}^<$ in Eq.~(\ref{eq:deu_kin}) can be identified. In the above equation,  the $1\pm f_i$ in the square bracket come from the quantum statistics of fermions ($-$) and bosons ($+$), and the $\delta$-function  denotes the conservation of energy and momentum with $p_\text{in}$ $=$ $\sum_{i=1'}^{3'}p_i$ and $p_{\rm out}$ $=$ $p_\pi +p_d$. The factor $3/4$ in the   third  line comes from the spin factors of initial and final states. The second line of Eq.~(\ref{eq:deu_rate}) denotes the spin-averaged squared amplitude at a relative distance ${\bf r}$ between the two nucleons.   The $W_d$ denotes the deuteron Wigner function,   with $\tilde{\bf r}$ $=$ $\tilde{\bf r}_n-\tilde{\bf r}_p$ and $\tilde{\bf p}$ $=$ $(\tilde{\bf p}_n-\tilde{\bf p}_p)/2$ being the relative  coordinate  and momentum in deuteron center-of-mass frame, respectively. For simplicity, we take $W_d$ $=$ $8e^{-\tilde{\bf r}^2/\sigma_d^2-\tilde{\bf p}^2\sigma_d^2}$ with $\sigma_d$ $=$ $3.2~\rm fm$ to reproduce the deuteron root-mean-square radius of $1.96~\rm fm$~\cite{SchPRC59}. 

For an approximately uniform system, the spatial part of $W_d$ in Eq.~(\ref{eq:deu_rate}) can be  integrated out, i.e., $|\phi_d(\tilde{\bf p})|^2$ $=$ $\int \text{d}^3 {\bf r}  \gamma_d W_d$ $=$ $(4\pi\sigma_d^2)^{3/2}e^{-\tilde{\bf p}^2\sigma_d^2}$ with $\gamma_d$ $=$ $E_d/m_d$, simplifying the second line of Eq.~(\ref{eq:deu_rate}) to  
$ 2m_d |\phi_d|^2\big{(}\overline{|\mathcal{M}_{\pi^+ n\rightarrow \pi^+ n}|^2}+ n\rightarrow p\big{)}$, which is the usual impulse approximation for $\overline{|\mathcal{M}_{\pi^+ d\rightarrow \pi^+ np}|^2}$, e.g. adopted in Ref.~\cite{DanNPA533} by treating deuteron as a point  particle.  
 
\begin{figure}[t]
  \centering
  \includegraphics[width=0.3\textwidth, bb=110 20 790 540]{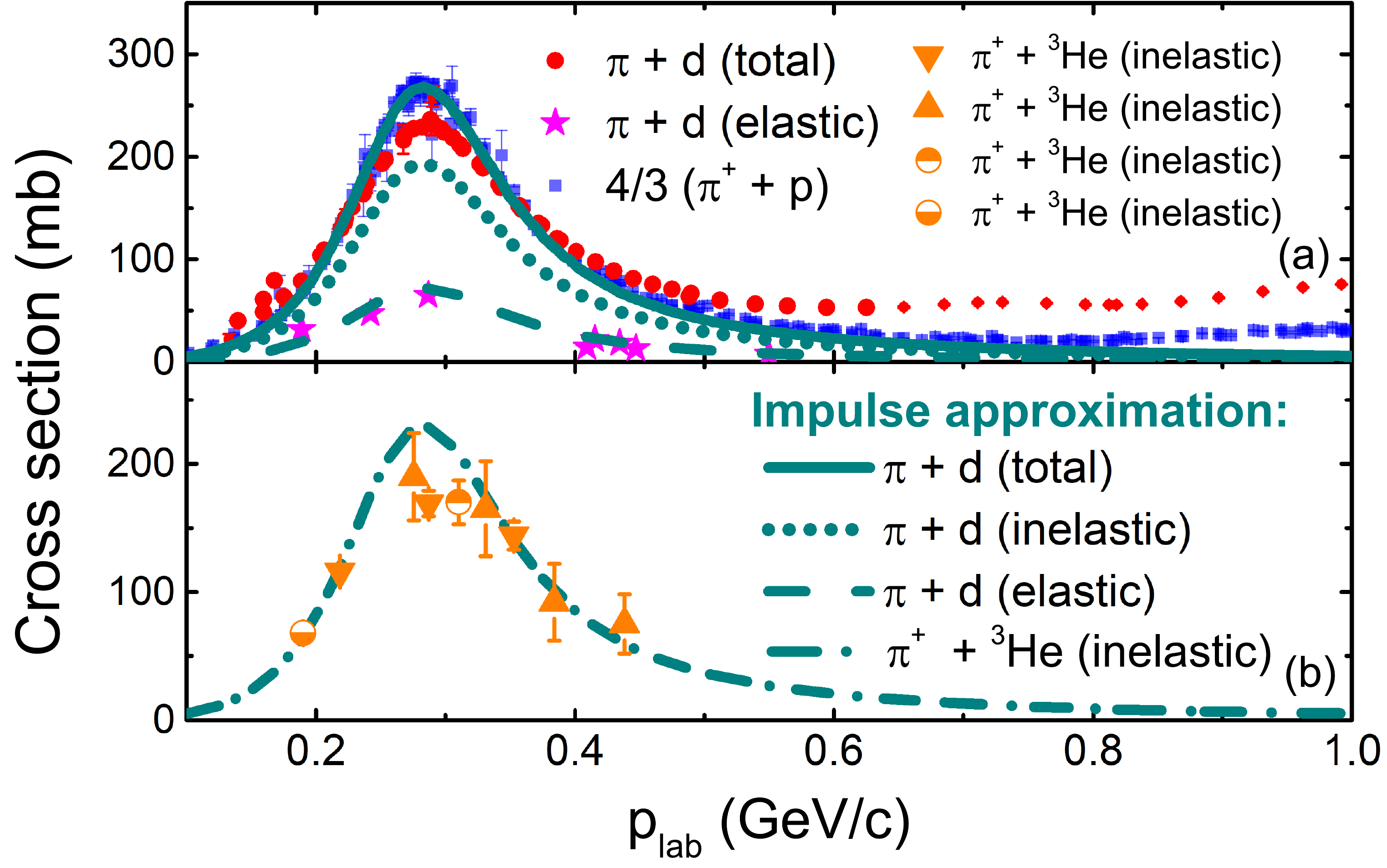}
 \caption{  Comparison of the pion incident momentum ($p_\text{lab}$) dependence  of the cross sections  for the reactions  $\pi +d$ (upper panel) and $\pi + ^3$He (lower panel)  between experimental data~\cite{WhiNPA408,KlePLB187,KleNPA472,KhaPRC44,ArnPRC50,YulPRC55,PDG2020} and our theoretical fit based on the impulsive approximation (colored lines).   }
  \label{pic:cs}
\end{figure}

The spin averaged squared pion-nucleon scattering  matrix element  $\overline{|\mathcal{M}_{\pi^+ n\rightarrow \pi^+ n}|^2}$ can  be related to the $\pi N$ scattering cross section. Under the impulse approximation, the  deuteron  dissociation cross section by a pion of momentum $p_{\rm lab}$ is approximately given by $\sigma_{\pi^+ d\rightarrow \pi^+ np}$ $\approx$ $\sigma_{\pi^+ n\rightarrow \pi^+ n}$ $+$ $\sigma_{\pi^+ p\rightarrow \pi^+ p}$. This relation becomes exact only for extremely large $p_\text{lab}$~\cite{DanNPA533,PDG2020}.
At low  $p_\text{lab}$, e.g. $0.3~\rm GeV$, one can introduce  a  renormaliztion factor $F_d$~\cite{DanNPA533} such that $\sigma_{\pi^+ d\rightarrow \pi^+ np}$ $=$ $F_d(\sigma_{\pi^+ n\rightarrow \pi^+ n}+\sigma_{\pi^+ p\rightarrow \pi^+ p})$.  As shown in Fig.~\ref{pic:cs}, using the constant values $F_d$ $\approx$ $0.72$ and $F_{\isotope[3]{He}}$ $\approx$ $0.51$ leads to an excellent description of the data for the $\pi$ $+$ $d$ and $\pi$ $+$ $\isotope[3]{He}$ cross sections in the energy region relevant for present study.  

Given the above deuteron dissociation and production cross sections with a pion, the probability for  the reaction $\pi^+d$ $\rightarrow$ $\pi^+np$ between  a pion and a deuteron inside a volume $\Delta V$ to  take place within a time interval $\Delta t$ can be obtained as 
\begin{eqnarray}
P_{23}\big{|}_\text{IA} \approx F_d{ v}_{12}\sigma_{\pi^+ p\rightarrow \pi^+ p}\frac{\Delta t}{{\Delta} V} + (2\leftrightarrow 3)\label{eq:deu_p23},
\end{eqnarray}
where ${{v_{12}}}$ is the relative velocity between the pion and one of the two nucleons inside deuteron, and the two terms on r.h.s correspond to the two diagrams in Fig.~\ref{pic:piond}.
Similarly, the probability for the reaction  $\pi^+np \rightarrow \pi^+d$ is 
\begin{eqnarray}
P_{32}\big{|}_\text{IA} \approx \frac{3}{4}F_d{ v}_{1'2'}\sigma_{\pi^+ p\rightarrow \pi^+ p}\frac{\Delta t}{\Delta V} W_d+(2'\leftrightarrow 3')\label{eq:deu_p32_wig}.  
\end{eqnarray} 
In obtaining Eqs. (\ref{eq:deu_p23}) and (\ref{eq:deu_p32_wig}), we have neglected the factors $1\pm f_i$ in Eq.~(\ref{eq:deu_rate}), which is a good approximation in high-energy nuclear collisions.
%\sout{With Eqs.~(\ref{eq:deu_p23}) and (\ref{eq:deu_p32_wig}), one can numerically solve Eq.~($\ref{eq:deu_kin}$) for deuteron production in transport model simulations of nuclear collisions. When the density gradient of the nucleons  is small }
Note that the deuteron Wigner function $W_d$ in Eq.~(\ref{eq:deu_p32_wig}) depends on both the coordinates and momenta of the constituent nucleons.
It should be replaced by $|\phi_d|^2/(\gamma_d \Delta V)$ if the deuteron is treated as a point particle.

In the limit that the  production and dissociation  rates are equal  in Eq.~(\ref{eq:deu_rate}), one obtains the  equilibrated  deuteron abundance  as
\begin{eqnarray}
N_d \approx \frac{3}{4}\int d\Gamma_{np} g_nf_n({\bf r}_n,{\bf p^*}_n)g_pf_p({\bf r}_p,{\bf p}_p)W_d(\tilde{\bf r},\tilde{\bf p}), \label{eq:deu_coallimit}
\end{eqnarray}
where $d\Gamma_{np}$ $=$ $(2\pi)^{-6}\text{d}^3{\bf r}_n\text{d}^3{\bf p}_n\text{d}^3{\bf r}_p\text{d}^3{\bf p}_p$  and the     neutron is chosen to be off mass shell to conserve energy~\cite{SchPRC59}. The   equilibrium  solution  given by Eq.~(\ref{eq:deu_coallimit}) is similar to the result from the phase-space coalescence model based on the   sudden approximation~\cite{BonPLB71}.  Note that the two nucleons in Eq.~(\ref{eq:deu_coallimit}) can be   $nn$, $np$, or $pp$ pair, while only the $np$ pair is considered in  the coalescence model.   For a uniform and thermalized system, Eq.~(\ref{eq:deu_coallimit}) can be further simplified to $N_d$ $\approx$ $g_dV(2\pi)^{-3}\int  \text{d}^3{\bf p}_d  f_n(\frac{{\bf p}_d}{2})f_p(\frac{{\bf p}_d}{2})$, which is the same as the deuteron abundance in chemical equilibrium with nucleons,  indicating that  our approach has the correct thermal limit.

\section{Numerical implementation and validation}
To solve Eq.~(\ref{eq:deu_kin}) with the collision integral given by Eq.~(\ref{eq:deu_rate}), we adopt the test particle ansatz ~\cite{WonPRC25} (or the particle in cell method), i.e., mimicking the distribution function $f_{\alpha}$ of a certain particle species of number $N_\alpha$ by a large number of delta functions, $f_{\alpha}({\bf r},{\bf p})$ $\approx$ $\sum^{N_{\alpha}N_{test}}_{i=1}\delta({\bf r}_i - {\bf r})\delta({\bf p}_i - {\bf p})$. Using test particles modifies Eq.~(\ref{eq:deu_p23}) and Eq.~(\ref{eq:deu_p32_wig}) by the multiplication of  factors  $1/N_\text{test}$ and   $1/N_\text{test}^2$, respectively. The number of test particles $N_\text{test}$ should be sufficiently large to ensure the convergence of the numerical results. To illustrate the numerical algorithm used in the present study, we take the reaction  $\pi d\rightarrow \pi NN$ as an example.   In each time step $\Delta t$,   a (test) pion and a (test) deuteron are randomly chosen from  a cell of volume ${\Delta}V$, and  the momenta of two nucleons inside the deuteron is then sampled according to its wave function with one nucleon  slightly off mass shell  to conserve energy and momentum.  After the scattering of the pion with the off-shell nucleon, both  final state nucleon and pion are taken to be on mass shell.  A similar consideration applies to the inverse reaction $\pi NN\rightarrow \pi d$.  One first randomly chooses a (test) pion and a (test) nucleon  during a time step $\Delta t$ from a cell of volume ${\Delta}V$. After letting the pion to scatter with the nucleon, their final momenta are then  sampled according to the differential cross section but with  the nucleon to be off-shell, which subsequently coalesces with the other nearby nucleon to form a deuteron.  Since we employ the stochastic method~\cite{DanNPA533,XZPRC71,WRFiP8},    the probabilities of these two processes  are directly evaluated  according to Eq.~(\ref{eq:deu_p23}) and Eq.~(\ref{eq:deu_p32_wig}), respectively.

As for triton, it can be formed from the two reactions $\pi NNN\leftrightarrow \pi t$ and $\pi dN\leftrightarrow \pi t$. The $3\leftrightarrow 2$ process   can be similarly treated as  for  deuteron production.   The  probability of the $4\leftrightarrow 2$ process  to occur in a volume $\Delta V$  during a time step $\Delta t$ is given by 
\begin{eqnarray}
P_{42}\big{|}_\text{IA}&\approx& \frac{1}{4}F_t\frac{{v}_{\pi N}\sigma_{\pi N \to\pi N}\Delta t}{N_\text{test}^3\Delta V} W_t, \label{eq:tri_p42}
\end{eqnarray}
where $F_t\approx F_{\isotope[3]{He}} \approx 0.51$ and the $W_t$ is the triton Wigner function, and its expression can be found in Refs.~\cite{SchPRC59,CLWNPA729}.
The branching ratio for the  dissociation of triton via the two inverse reactions can be  estimated by calculating the deuteron component in the triton wave function.
If we take the wave function to have the Gaussian form, we have the branching ratio $\mathcal{B}(\pi t\rightarrow \pi Nd)  \approx 0.56$.
More realistic Hulthen wave function for the deuteron leads to a similar branching ratio $\mathcal{B}(\pi t\rightarrow \pi Nd)  \approx 0.58$
Similarly, the branching ratio for $^3$He dissociation to a deuteron is 
$\mathcal{B}(\pi ~^3\text{He}\rightarrow \pi Nd) \approx 0.6$.
We note the following numerical results are insensitive to the value of the branching ratios in the  production and dissociation of triton and helium-3.

\begin{figure*}[t]
  \centering
 \includegraphics[width=17cm]{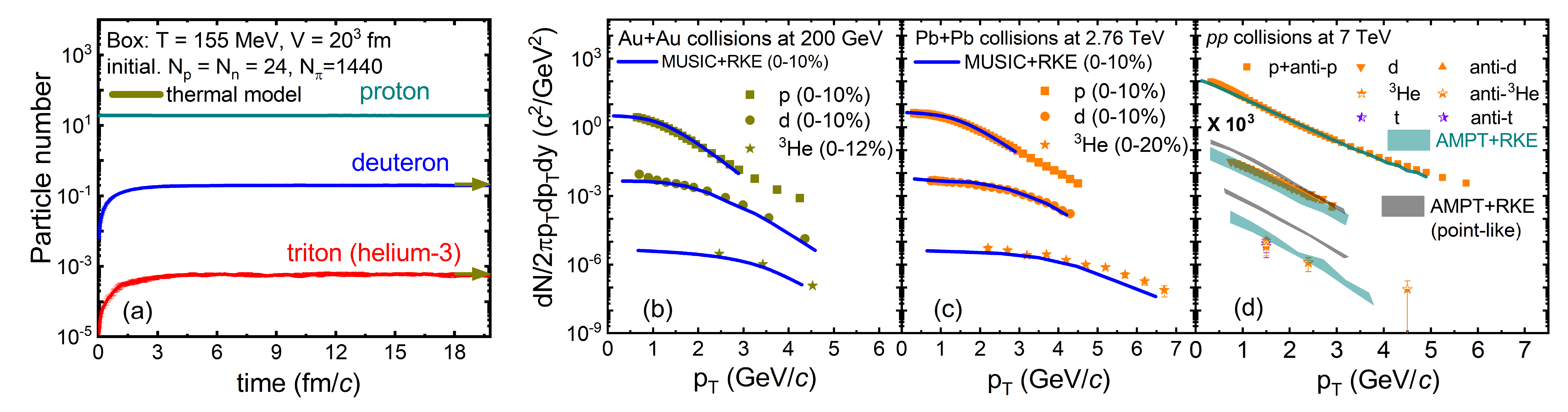}
  \caption{Panel (a) Box calculations for deuteron and triton (helium-3). Transverse momentum spectra of $p$, $d$, and $^3$He ($t$)  in (b) central Au+Au collisions at $\sqrt{s_{\rm NN}}=200~\rm GeV$; (c) central Pb+Pb collisions at $\sqrt{s_{\rm NN}}=2.76~\rm TeV$; (d) $pp$ collisions at $\sqrt{s_{\rm NN}}=7~\rm TeV$. The experimental data points are taken from Refs.~\cite{STARPRL97,STAR0909,ALICEPRC88,ALICEPRC93,STARPRC99,ALICEEPJC75,ALICEPRC97}. Theoretical predictions from   the relativistic kinetic equations (RKE) are denoted by colored lines. The spectra in panel (d) are multiplied by a factor of 10$^3$.}
  \label{pic:RQKE}
\end{figure*}

To validate  the above described algorithm, we  consider deuteron and triton production from a hadronic matter in   a (20 fm)$^3$ box with periodic boundary conditions. The box consists of 24 protons, 24 neutrons, and 480 pions for each of its three charge states. The initial distribution of these particles is taken to be uniform in the coordinate space   and to have  a thermal Boltzmann distribution with temperature $T = 155~\rm MeV$  in the momentum space.  Fig.~\ref{pic:RQKE} (a) shows the time evolution of the yields of proton, deuteron, and triton (helium-3).   It is seen that  their final numbers   are consistent with their thermal values at chemical equilibrium  over 5 orders of magnitude.

\section{Light nuclei production in high-energy nuclear collisions}
We now apply the above described kinetic  approach  to different  collision systems,  i.e., central Au+Au collisions at $\sqrt{s_{\rm NN}}=200~\rm GeV$, Pb+Pb collisions at $\sqrt{s_{\rm NN}}=2.76~\rm TeV$, and $pp$ collisions at $7~\rm TeV$. For  the evolution of the QGP produced in central Au+Au and Pb+Pb collisions,   we use  the (3+1)-d viscous hydrodynamic  model  MUSIC~\cite{PaqPRC93,SCPRC97,DenPRC98,SCNST31} with   initial conditions taken from a newly developed dynamical Glauber model~\cite{SCPRC97,SCNPA982}. The QGP has a smooth crossover  to a hadronic matter along a phase transition line $T(\mu_B)$ described in Ref.~\cite{MonPRC100}.  The initial hadrons are sampled on a constant energy density particalization hypersurface according to  the Cooper-Frye formula~\cite{CooPRD10}.  For $pp$ collisions,  the  initial hadron distribution is generated from a multiphase transport  (AMPT) model~\cite{LZWPRC72}. In the present article, we also assume that   light nuclei are initially produced from hadronization of QGP,  and their yields are estimated from the canonical statistical model~\cite{VovPLB785}  that conserves baryon, electric and strange charges.

For the evolution of the hadronic matter and the production of light nuclei,  besides the production and dissociation of deuteron and helium-3 by pions, most scattering channels in AMPT~\cite{LZWPRC72} have   been included. These channels describe properly the dynamics of pions and nucleons   that are relevant for   light nuclei production from the reactions discussed  above. 

Fig.~\ref{pic:RQKE} (b) and (c) show the comparison of  the transverse momentum ($p_T)$ spectra of $p$, $d$, and $^3$He between our predictions and the experiment data~\cite{STARPRL97,STAR0909,ALICEPRC88,ALICEPRC93,STARPRC99} in central Au+Au (Pb+Pb) collisions at $\sqrt{s_{\rm NN}} = 200~\rm GeV$~(2.76~TeV). The solid lines denote predictions from  relativistic kinetic equations with the inclusion of the finite sizes of $d$ and $\isotope[3]{He}$, and they describe  the data very well.  We have also  studied the case when treating light nuclei as point particles, and the results remain in agreement with experiment data within uncertainties. This is not surprising as the sizes of the produced matter in these collisions are over $10~\rm fm$  and are much larger than those of $d$ and $\isotope[3]{He}$. Besides, we find the final yields of $d$ and $\isotope[3]{He}$ are insensitive to their initial values at QGP hadronization, similar to that found  in Refs.~\cite{OhPRC80,OliPRC99} for the deuteron. The information of initial light nuclei is largely lost during the hadronic evolution which lasts for more than $15~{\rm fm}/c$.

The situation, however, changes dramatically in  $pp$ collisions at $\sqrt{s_{\rm NN}} = 7~\rm TeV$ as shown  in  Fig.~\ref{pic:RQKE} (d). For the initial $d$ and $\isotope[3]{He}$ generated from hadronization, there are uncertainties  of the canonical effects~\cite{VovPLB785} for $pp$ collisions, which are absent in large systems. The initial $d$ and $\isotope[3]{He}$ in central rapidity ranges approximately from $1.3\times10^{-4}$ to $5.3\times10^{-4}$, and from $7.3\times10^{-8}$ to $1.3\times10^{-6}$, respectively. The lower values correspond to an aggressive estimation of the canonical effects, while the upper  ones  are obtained by neglecting the canonical effects.  The uncertainty in our theoretical predictions on the $p_T$ spectra of $d$ and \isotope[3]{He}, shown in  Fig.~\ref{pic:RQKE} (d) by the shaded bands, reflects the uncertainty in the initial $d$ and $\isotope[3]{He}$ yields. For the case of treating light nuclei  as point particles, we find the spectra of $d$ and \isotope[3]{He} are significantly overestimated, regardless of  the initial yields of $d$ and \isotope[3]{He}. After taking into account of their sizes, both spectra are  reduced, and they now agree  with the experimental data.  Unlike in central Au+Au and Pb+Pb collisions, we find the final yields of $d$ and \isotope[3]{He} in $pp$ collisions to be  sensitive to their initial values  after their sizes are taken into account. This is  mainly because  the reaction rate  is not large enough to achieve chemical equilibrium among nucleons, deuterons and helium-3 during the very short collision time of  merely a few fm/$c$. As a result, both the hadronization mechanism and the many-body hadronic reactions play important roles in light (anti-)nuclei production in high-energy $pp$ collisions. This finding will  also be relevant for the indirect  search of dark matter from studying light (anti-)nuclei in cosmic rays.

\section{Summary}
We  have developed a stochastic method to solve the relativistic kinetic equations for light nuclei production from many-body reactions in the hadronic matter produced in high-energy nuclear collisions.  With the initial hadron distributions generated from the MUSIC  hydrodynamic model or the AMPT model, we have found that  the $d$ and $^3$He $p_T$ spectra measured  in central Au+Au (Pb+Pb) collisions at $\sqrt{s_{\rm NN}}=200~\rm GeV$ ($2.76~\rm TeV$) and in $pp$ collisions at $7~\rm TeV$ are well described by the present approach. In particular, the inclusion of light nuclei sizes, which has been neglected in previous transport studies, is essential for describing their suppressed production  in $pp$ collisions.  The present kinetic approach has thus significantly advanced our understanding of light nuclei production in high-energy nuclear collisions. This is different from the usual  statistical hadronization model and the nucleon phase-space coalescence model, which correspond to the limits of our dynamic model if the final hadronic matter in high-energy nuclear collisions is a uniform and thermalized system.

The present study can be extended straightforwardly to investigate the production of other light nuclei like $\isotope[4]{He}$ (\isotope[4]{Li}), $^3_\Lambda$H, $^4_\Lambda$H, and exotic states like  $X$(3872)~\cite{BellePRL91} and the possible $\Omega$-dibaryon~\cite{ExHICPRC84,ClePPNP93,ExHICPPNP95}  in various collision systems, such as the $e\bar{e}$, $pp$, $pA$ and $AA$  collisions.  With the upcoming high-quality data of light nuclei from RHIC and LHC, the present kinetic approach will further help shed light on more fundamental questions in physics, such as cosmic-ray dark matter detection and the QCD phase structure in relativistic heavy-ion collisions.

\begin{acknowledgments}
\emph{Acknowledgments}{\bf ---}We thank  Lie-Wen Chen, Zi-Wei Lin, Feng Li,  Zhen Zhang, and Xiao-Jian Du for helpful discussions, and Chen Zhong for setting up and maintaining the GPU server. This work was supported in part by the U.S. Department of Energy under Award No. DE-SC0015266, No. DE-SC0013460, the Welch Foundation under Grant No. A-1358, the National Science Foundation (NSF) under grant number PHY-2012922, National Natural Science Foundation of China under contract No. 11891070 and No. 11890714, and the Guangdong Major Project of Basic and Applied Basic Research No. 2020B0301030008.
\end{acknowledgments}

\bibliography{RKE-LN}

\end{document}